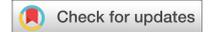

# Kondo-like behavior in a mixed valent oxypnictide $La_3Cu_4P_4O_2$

Szymon Królak[1✉], Michał J. Winiarski[1], Duygu Yazici[1,2], Soohyeon Shin[3] & Tomasz Klimczuk[1✉]

We have synthesized and characterized the physical properties of a layered, mixed valent oxypnictide $La_3Cu_4P_4O_2$ via magnetization, electrical resistivity, and specific heat measurements. Although $La_3Cu_4P_4O_2$ does not exhibit superconductivity down to T = 0.5 K, it demonstrates an intriguing resistivity minimum observed at $T_{min}$ = 13.7 K. Disappearance of the resistivity minimum under an applied magnetic field of $\mu_0 H$ = 9 T together with the negative magnetoresistance at low and positive at high temperatures are observed, which are typical for both Kondo-like spin-dependent scattering and 3D weak localization. We argue that the Kondo scattering is a more plausible explanation due to the low-temperature deviation from the Curie-Weiss law observed in the magnetic susceptibility, consistent with the presence of magnetic interactions between paramagnetic $Cu^{2+}$ ions and Kondo screening of these $Cu^{2+}$ moments. We supplemented the experimental characterization with a detailed description of chemical bonding, employing density functional theory (DFT) calculations and crystal orbital Hamilton population (COHP) analysis for $La_3Cu_4P_4O_2$ and isostructural $La_3Ni_4P_4O_2$, which is a superconductor with $T_c = 2.2$ K. Based on the calculations performed, we present the difference between $La_3Cu_4P_4O_2$ and $La_3Ni_4P_4O_2$ in the character of electronic states at the Fermi level. This discrepancy impacts structural stability and may cause a lack of superconductivity in $La_3Cu_4P_4O_2$ down to T = 0.5 K.

One of the most active fields in condensed matter physics is the study of superconductivity. In particular, intensive research has been performed on layered transition-metal pnictides (Pn = As, P) since the discovery of high-temperature superconductivity in Fe-pnictides[1,2] and chalcogenides in 2008[3], with critical temperatures reaching as high as 56 K in $Gd_{1-x}Th_xFeAsO$[4] and 100 K in single-layer FeSe films grown on doped $SrTiO_3$[5]. The 3442 family, $Ln_3T_4Pn_4O_2$ (Ln = lanthanide; T = Ni, Cu; Pn = P, As), while characterized by much lower superconducting transition temperatures, is nevertheless interesting. Its crystal structure can be viewed as a convolution of the $ThCr_2Si_2$ and ZrCuSiAs structures. There are two known superconductors in the 3442 family: $La_3Ni_4As_4O_{2-\delta}$ and $La_3Ni_4P_4O_2$, with critical temperatures $T_c$ = 2.0 K and $T_c$ = 2.2 K, respectively[6,7]. The stacking of inequivalent charge reservoir layers, $(La_2O_2)^{2+}$ from one side and $La^{3+}$ from the other, leads to an asymmetric charge distribution around the $Ni_2P_2$ planes, a feature that distinguishes $La_3Ni_4Pn_4O_2$ (Pn = P, As) from the other known superconducting oxypnictides. To understand the relation between the crystal structure and superconductivity in the 3442 family, the physical properties of the $La_3Cu_4P_4O_2$, together with the electronic structure of both $La_3Cu_4P_4O_2$ and $La_3Ni_4P_4O_2$ were investigated in the present paper. $La_3Cu_4P_4O_2$ was first synthesized by Cava et al. in 1997[8], showing no signs of magnetic ordering above T = 4.2 K and a lack of superconductivity down to T = 2 K.

Besides superconductivity, resistivity minima in various systems-such as metallic alloys, metallic glasses, ceramics, and epitaxially grown thin films-has also been a topic of considerable interest for years[9–18]. Extensive experimental and theoretical studies have been conducted to investigate the origin of such behavior, with several mechanisms proposed. These include e.g. spin-polarized tunneling via grain boundaries[14], Kondo-like effects[9,18], and quantum interference effects arising from electron-electron interactions (EEI) and weak localization (WL)[14,17]. Our electrical resistivity measurements of $La_3Cu_4P_4O_2$ indicate lack of superconductivity down to T = 0.5 K, rather unexpectedly showing a resistivity minimum ($\rho_{min}$) at T = 13.7 K. Therefore, in this paper, besides the magnetic, electronic and thermal properties of $La_3Cu_4P_4O_2$, possible mechanisms behind the resistivity minimum have been considered. Our results indicate that the origin of the low-temperature resistivity minimum behavior is presumably due to the Kondo scattering by magnetic moments of $Cu^{2+}$. However, WL could not be ruled out due to the measurement constraints.

[1]Faculty of Applied Physics and Mathematics and Advanced Materials Centre, Gdansk University of Technology, ul. Narutowicza 11/12, 80-233 Gdańsk, Poland. [2]The Scientific and Technological Research Council of Turkey, Atatürk Bulvarı No: 221, Kavaklıdere 06100, Ankara, Turkey. [3]PSI Center for Neutron and Muon Sciences, Paul Scherrer Institut, 5232 Villigen PSI, Switzerland. ✉email: szymon.krolak@pg.edu.pl; tomasz.klimczuk@pg.edu.pl

## Synthesis

Polycrystalline samples of $La_3Cu_4P_4O_2$, $La_3(Cu_3Ni)P_4O_2$, and $La_3Cu_4P_4O_{1.9}F_{0.1}$ were prepared with a solid-state reaction method from stoichiometric amounts of powdered $La_2O_3$ (Alfa Aesar, 99.9%, dried at 900°C overnight before use), $CuP_2$, Cu (Merck, analytical), Ni (Puratronic; 99.996%, metal basis), $LaF_3$ (Alfa Aesar, 99%), and La (Johnson M., 99.99%) chunks. The $CuP_2$ precursor was prepared by reacting Cu powder and red phosphorus (Alfa Aesar Puratronic, 99.999+%) in an evacuated quartz tube in two steps. First, the ampoule was heated slowly (20°C/h) to 600°C and annealed at this temperature for 25 hours, followed by quenching in air. Before the second firing, the tube was shaken in order to break the newly-formed grains and ensure that the reaction between Cu and P takes place not only at the surface but also in the volume. In the second step, the ampoule was heated from 600°C to 720°C at the rate of 180°C/h and kept at 720°C for 80 hours.

$La_3Cu_4P_4O_2$ and $La_3Cu_4P_4O_{1.9}F_{0.1}$ samples were synthesized as follows. Stoichiometric amounts of precursors were weighed, thoroughly mixed, pressed into pellets, sealed in evacuated quartz tube, and annealed with a three-step protocol: (1) at 1070°C for 45 hours, followed by grinding in an Ar-filled glovebox to achieve homogeneity, pressed into pellets again, and (2) annealed in evacuated quartz tube at 1100°C for 21 hours. The procedure was repeated one more time (3) by annealing the material at 1120°C for 66 hours.

For the $La_3(Cu_3Ni)P_4O_2$ sample, the initial preparation was the same, only with a different annealing procedure: (1) at 1070°C for 19 hours, (2) at 1170°C for 24 hours and (3) at 1180°C overnight, followed by quenching in water.

## Methods

Crystal structure and sample quality of the $La_3Cu_4P_4O_2$, $La_3Cu_4P_4O_{1.9}F_{0.1}$ and $La_3(Cu_3Ni)P_4O_2$ samples were characterized through powder XRD (pXRD) diffraction, performed using Bruker D2 Phaser 2nd generation diffractometer with $Cu_{K\alpha}$ radiation ($\lambda$ = 1.5404 Å) and XE-T position-sensitive detector. The obtained pXRD data were refined using the LeBail method as implemented in the FullProf Suite[19]. Magnetization, electrical resistivity, and heat capacity measurements were carried out with Quantum Design Physical Property Measurement System (PPMS) Evercool II in the temperature range T = 1.9 - 300 K. Electrical resistivity - $\rho$ (T) - measurements of $La_3Cu_4P_4O_2$ sample below T = 1.9 K were performed using a PPMS equipped with a He-3 insert. $\rho$(T) of the $La_3Cu_4P_4O_2$, $La_3Cu_4P_4O_{1.9}F_{0.1}$, and $La_3(Cu_3Ni)P_4O_2$ samples were determined using the four-terminal technique, with 50 $\mu$m-diameter platinum wires attached to the sample surface with the two-component paste (EPO-TEK H20E). Specific heat measurements were performed using a two-$\tau$ relaxation method, and throughout the text, the magnetic susceptibility is approximated as $\chi \approx M/H$.

Density functional theory (DFT) calculations of the electronic structure of $La_3Cu_4P_4O_2$ and $La_3Ni_4P_4O_2$ were performed using the Quantum ESPRESSO 7.2 (QE) package[20–22]. Projector augmented wave (PAW)[23,24] sets were taken from the PSlibrary[25]. Self-consistent field calculations were performed on a 13x13x4 k-point grid within the irreducible wedge of the BZ, while for crystal orbital Hamilton population (COHP) and density of states (DOS) analysis, a 22x22x7 k-point grid was used. The Perdew-Burke-Ernzerhof Generalized Gradient approximation[26] was used for the exchange-correlation potential. COHP[27,28] analysis was performed by projecting the QE plane wave output to a local orbital basis by Koga et al.[29] using the Lobster 5.0.0 code[30,31].

## Results and discussion

### Crystal structure

Fig. 1. displays the powder X-ray diffraction pattern of $La_3Cu_4P_4O_2$. We confirm that $La_3Cu_4P_4O_2$ crystallizes in the the ordered variant of the $Zr_3Cu_4Si_6$-type structure (I4/mmm, s.g. no 139)[32] with lattice constants a = 4.040(1) Å and c = 26.821(3) Å, nearly identical to those reported earlier[8,33] and both larger than the lattice constants of $La_3Ni_4P_4O_2$: a = 4.0107(1) Å, c = 26.1811(9) Å[7]. Our X-ray results pointed to contamination by a single, nonmagnetic impurity phase $La_3PO_7$[34] which is indicated by an asterisk in Fig. 1.

In the inset of Fig. 1., the crystal structure of $La_3Cu_4P_4O_2$ is shown. The ordered variant of the $Zr_3Cu_4Si_6$-type structure can be viewed as an intergrowth of hypothetical LaCuPO and $LaCu_2P_2$ compounds, in which $La_2O_2$ and La layers separate the electronically active $Cu_2P_2$ layers. Each copper atom is surrounded by four P atoms forming a tetrahedron, which is the most common coordination environment for $Cu^+$[35]. Initially, Kaiser and Jeitschko[32] assumed that copper is fully monovalent, and proposed the formula $Ln_3Cu_4P_4O_{1.5}$ to obtain charge balance in the structure, underpinning their assertion of an anticipated semiconducting nature of the compounds. This approach was subsequently applied by Wang et al.[6], in their investigation of isostructural oxyarsenides. In their work, they employed the formulation $Ln_3M_4As_4O_{2-x}$, despite the observed metallic behavior of the samples and the absence of empirical evidence supporting the existence of oxygen vacancies. In a subsequent paper,[36] the authors performed neutron diffraction measurements and concluded that the oxygen position is fully occupied, rendering the studied compound stoichiometric with the formula $Ce_3Cu_4As_4O_2$. Similarly, in a theoretical scenario where all copper atoms exclusively adopt the $Cu^+$ oxidation state, $La_3Cu_4P_4O_{1.5}$ would be composed of closed-shell ions only, thereby implying a semiconducting nature. However, contrary to this expectation, our result does not indicate such behavior. Moreover, it was shown that there is an equal amount of $P^{3-}$ and $P^{2-}$ in the $La_3Cu_4P_4O_2$ and that the oxygen occupancy factor is close to 1[33]. The presence of $P^{2-}$ species results from the dimerization of 2 out of 4 P atoms per formula unit, confirmed by electronic structure calculations presented here. Thus, the only way to obtain a net charge of zero is to include one $Cu^{2+}$ per f. u., according to the formula: $(La_3)^{9+}(Cu_3)^{3+}(Cu)^{2+}(P_2)^{6-}(P_2)^{4-}(O_2)^{4-}$. It is, however, difficult to anticipate ionic Cu-P bonds; thus, following[33], we reformulate the composition as $(La_3)^{9+}(Cu_4P_4)^{5-}(O_2)^{4-}$, where the electronic adjustment occurs within the $(Cu_4P_4)^{5-}$ anion. Two degrees of freedom govern the electron distribution in this anion: (1) the oxidation state of Cu and (2) the P-P bond length. The P-P bonds demonstrate considerable flexibility, either shortening to decrease the negative charge on phosphorus (which requires fewer $Cu^{2+}$ ions for charge compensation) or elongating to increase the negative charge (which necessitates more $Cu^{2+}$ ions). Given that the phosphorus atoms cannot possess a formal charge more negative than 3-, and that half of the phosphorus atoms are dimerized, it is essential that some amount of $Cu^{2+}$ ions, hybridized with surrounding P atoms, exist within the studied material. It is also noteworthy that Cu atoms occupy only one symmetry-inequivalent site in $La_3Cu_4P_4O_2$, implying that the proposed mixed valence cannot exhibit a localized character.

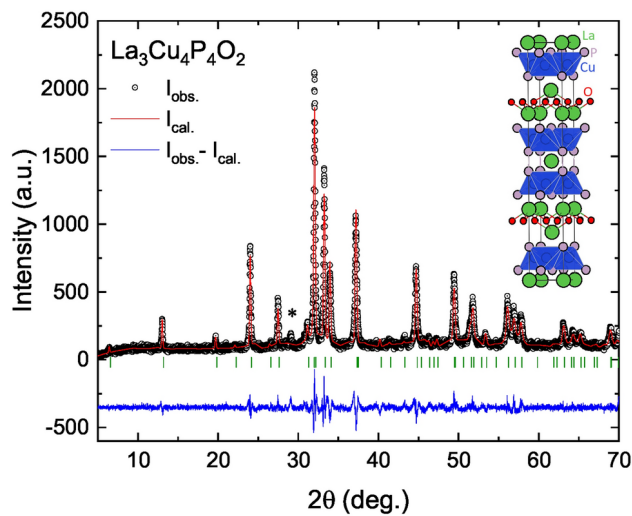

**Fig. 1**. $La_3Cu_4P_4O_2$ powder diffraction pattern with a unit cell presented in the inset. Experimental data points (white) were refined using the LeBail algorithm implemented in FullProf Suite. The obtained fitting profile (red line) matches well with the observed pattern. The difference between observed and calculated intensities is presented in the blue line, whereas positions of diffraction peaks are denoted as green dashes. The only visible impurity peak is marked with an asterisk.

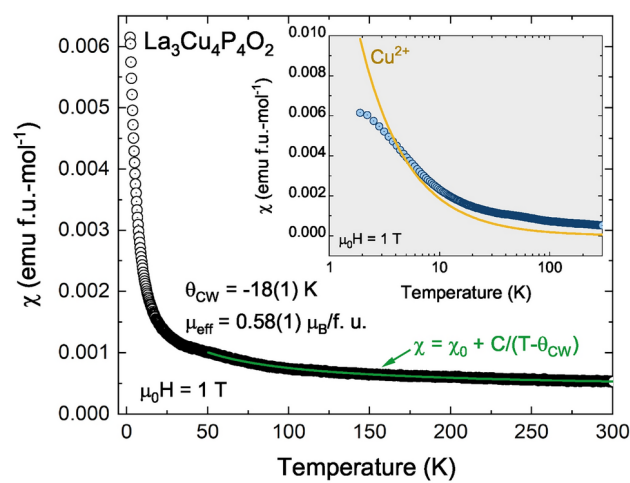

**Fig. 2**. Magnetic susceptibility data per formula unit of $La_3Cu_4P_4O_2$ measured in the applied field of 1 T, with the diamagnetic contribution of the sample holder subtracted. The data were modeled with the modified Curie-Weiss law in the temperature range 50 K - 300 K (green line). The inset shows the same data vs log T, emphasizing the departure from the localized model below approx. 10 K.

### Magnetic properties
Temperature dependence of the magnetic susceptibility ($\chi$), measured in an applied field of 1 T is presented in Fig. 2., with the diamagnetic contribution of the sample holder subtracted (inverse susceptibility plot is presented in SI Fig. S1.). The susceptibility shows no sign of magnetic order down to T = 1.9 K and looks similar to what is expected for the presence of localized magnetic moments, however, the low-temperature upturn is rather weak, resembling the so-called "Curie tail". Such a behavior is usually ascribed to either magnetic defects or impurity phases[37–39] present in the sample, however, it was recently shown that the mechanism can be intrinsic[40]. The observed upturn is in agreement with the presence of localized $Cu^{2+}$ ions in the sample needed to achieve charge balance as mentioned earlier, although as shown in the inset of Fig. 2. there is a visible discrepancy between the susceptibility of a localized $Cu^{2+}$ ion and the data for $La_3Cu_4P_4O_2$ below T = 10 K. It might be caused by the onset of magnetic correlation, i.e. short-range magnetic order mediated by an RKKY interaction between the Cu ions, combined with a reduction in the effective magnetic moment of the Cu ions due to the Kondo screening by the conduction electrons.

To extract the values of the Curie-Weiss temperature $\theta_{CW}$ and the effective magnetic moment $\mu_{eff}$, we performed modified Curie-Weiss law[41] fitting in the 50 - 300 K temperature range, according to the formula (Eq. 1): $\chi(T) = \chi_0 + \frac{C}{T-\theta_{CW}}$, where $\chi_0 = 3.98(3) \cdot 10^{-4}$ emu is the temperature-independent (Pauli paramagnetic) contribution. The effective moment $\mu_{eff}$ was extracted from the Curie constant, $\mu_{eff} = \left(\frac{3Ck_B}{N_A\mu_B^2}\right)^{\frac{1}{2}}$, where $N_A$ is Avogadro's number and $k_B$ Boltzmann's constant. This analysis was performed by fitting Eq. 1 to the data using non-linear least squares regression. The resulting best-fit parameter values for $\mu_{eff}$ and $\theta_{CW}$ are 0.58(1) $\mu_B$/f.u. and -18(1) K, respectively. Such a small value of $\mu_{eff}$, as compared to $\mu_{eff} = 1.73 \mu_B$ for a free $Cu^{2+}$ ion is similar to the value found in Ru-rich (x ≥ 3) $CaCu_3Ti_{4-x}Ru_xO_{12}$[42,43] and suggests that there is a small amount of $Cu^{2+}$ present in the sample, i.e. the valence of copper ions is mixed, between 1+ and 2+. To estimate the amount of $Cu^{2+}$, we performed the same fit but keeping the Curie constant fixed, fitting the fraction of the magnetic ions (see SI Fig. S2. for details). The obtained fraction n = 9.4(1) % means that almost 10 % of copper atoms per f.u. in $La_3Cu_4P_4O_2$ are in the 2+ oxidation state. This means that each Cu atom in the structure is in 2.5% in the magnetic, $Cu^{2+}$ state.

The magnetic susceptibility of the studied sample is comparable to that of $La_5Cu_4P_4O_4Cl_2$[33], a distinct compound that shares the same $(Cu_4P_4)^{5-}$ anion, making the comparison valid. In both materials, there is a low-temperature upturn in magnetic susceptibility on the order of $10^{-3}$ emu f.u.-mol$^{-1}$. However, the upturn observed in the $La_3Cu_4P_4O_2$ sample studied in[8] is an order of magnitude smaller. While this difference could be attributed to better sample quality (i.e., fewer $Cu^{2+}$ ions from impurities) in[8], we propose an alternative explanation. Both[33] and our study focus on samples synthesized with stoichiometric amounts of the reagents. In contrast,[8] reports that their best samples were prepared using an excess of phosphorus, corresponding to nominal compositions of $La_3Cu_4P_{4.05}O_2$ and $La_3Cu_4P_{4.1}O_2$. The direct effect of this excess phosphorus on the structural properties remains unclear, as neither[8] nor[33] provide powder X-ray diffraction (pXRD) data. In[33], the authors note that their sample is "single phase, while in[8] the samples are labeled as "best". Our sample exhibits lattice constants of a = 4.040(1) Å and c = 26.821(3) Å, which are nearly identical to those reported in[33] (a = 4.0389(4) Å, c = 26.817(3) Å). These values deviate slightly from those reported in[8] (a = 4.033(1) Å, c = 26.765(8) Å). Thus, it is reasonable to conclude that our sample is more similar to that studied in[33], as corroborated by the magnetic susceptibility measurements. The variations in lattice constants are mirrored by changes in the P-P bond length. In[8], authors report a bond length of 2.23 Å, compared to 2.27 Å in[33]. This shorter bond in[8] suggests less negative charge on the phosphorus atom, reducing the number of $Cu^{2+}$ ions required for charge compensation. In contrast, the longer bond of[33] corresponds to a more negative charge on phosphorus and a higher proportion of $Cu^{2+}$ ions. Although the difference in bond length appears small, it is significant given that the difference between singly and doubly bonded phosphorus is approximately 0.2 Å[44], making the 0.04 Å variation non-negligible. Thus, we ascribe the variation in the $Cu^{2+}$ content between the samples studied in this work, in[33] and the one from[8] to different synthetic conditions, which may influence the P-P bond length and in effect the electronic distribution inside the $(Cu_4P_4)^{5-}$ anion.

## Specific heat

Specific heat vs temperature data in the 1.9-300 K temperature range are shown in Fig. 3. The data were described by an expression $C_p(T) = \gamma T + n_D C_{Debye} + n_E C_{Einstein}$, where the first term is the contribution from the electron gas, while $C_{Debye}$ and $C_{Einstein}$ are the Debye and Einstein contributions to the specific heat, respectively. Here, we observe that the sum of the contributions from the normal modes of lattice vibration denoted as $n_D + n_E$, does not equate to the theoretically predicted value of 13 ($n_D + n_E \neq 13$), evidencing that 300 K is still too low to reach the Dulong-Petit limit. Based on the fit, we determined the number of oscillators $n_D = 9.9(1)$ and $n_E = 1.9(1)$, with $n_D + n_E$ close to 12, as well as the values of the Sommerfeld coefficient $\gamma = 9.5(1)$ mJ mol$^{-1}$ K$^{-2}$, Debye temperature $\theta_D = 354(3)$ K and Einstein temperature $\theta_E = 106(1)$ K. The electronic, Debye and Einstein contributions to specific heat, characterized by $\gamma$, $\theta_D$, and $\theta_E$ respectively, were also determined from the low-temperature fit to the $C_p/T$ vs T data, as shown in the inset of Fig. 3. The experimental data were analyzed using the formula $C_p/T = \gamma + \beta T^2 + \frac{C_E}{T}$, where the Einstein heat capacity is given by $C_E = A \cdot R \cdot \left(\frac{\theta_E}{T}\right)^2 \exp\left(\frac{\theta_E}{T}\right) \left(\exp\left(\frac{\theta_E}{T}\right) - 1\right)^{-2}$. Here, R = 8314 mJ mol$^{-1}$ K$^{-1}$ represents the gas constant, and A = 3.5(2) is a prefactor[45]. The extracted parameters $\gamma = 7.5(6)$ mJ mol$^{-1}$ K$^{-2}$, Debye temperature $\theta_D = 357(4)$ K, and Einstein temperature $\theta_E = 103(2)$ K are consistent with those obtained from the fit over the broader temperature range of 1.9-300 K. The number of oscillators, $n_E = A/3 = 1.15(6)$, is approximately half the value obtained from the fit over the full range. This discrepancy is reasonable, as the maximum in $C_{lattice}/T^3$ occurs at T = 24 K, and the present fit is restricted to data up to approximately 18 K.

In order to verify the consistency of the obtained values, we have additionally determined the values of the Sommerfeld coefficient from the linear fit to the data plotted as $C_p/T$ vs. $T^2$ at low temperatures (below T = $\Theta_D/50$) and the value of the Einstein temperature from the $C_p/T^3$ vs. T plot (see SI figures S3 and S4, respectively). The obtained values of $\gamma = 11.8(1)$ mJ mol$^{-1}$ K$^{-2}$ and Einstein temperature $\theta_E = 121$ K exhibit consistency with the ones mentioned earlier and are also corroborated by the value derived from the electronic structure calculation $\gamma = 8.3$ mJ mol$^{-1}$ K$^{-2}$. Comparing the compounds $La_3Cu_4P_4O_2$ (Cu-3442) and $La_3Ni_4P_4O_2$ (Ni-3442), it is observed that the parameter $\gamma$ decreases significantly in $La_3Cu_4P_4O_2$ compared to $La_3Ni_4P_4O_2$. Specifically, it is $\gamma \approx 1.9 - 3.0$ mJ mol$^{-1}$ K$^{-2}$/Cu in Cu-3442, whereas $\gamma \approx 6.2$ mJ mol$^{-1}$ K$^{-2}$/Ni in Ni-3442. Furthermore, while the theoretical and experimental values of $\gamma$ for Cu-3442 align closely, in the case of Ni-3442, the theoretical value $\gamma = 3.1$ mJ mol$^{-1}$ K$^{-2}$/Ni obtained in this study is only half of the experimental one. This suggests the presence of interactions in Ni-3442 that alter the Sommerfeld coefficient beyond its value for free electrons. As the value of $\gamma$ is proportional to the density of states at the Fermi level, the above comparison points to the fact that there is less d-electron density of states at the Fermi level - DOS($E_F$) - in the Cu-bearing compound, which conclusion is substantiated by the electronic structure calculations presented in the later part of this paper.

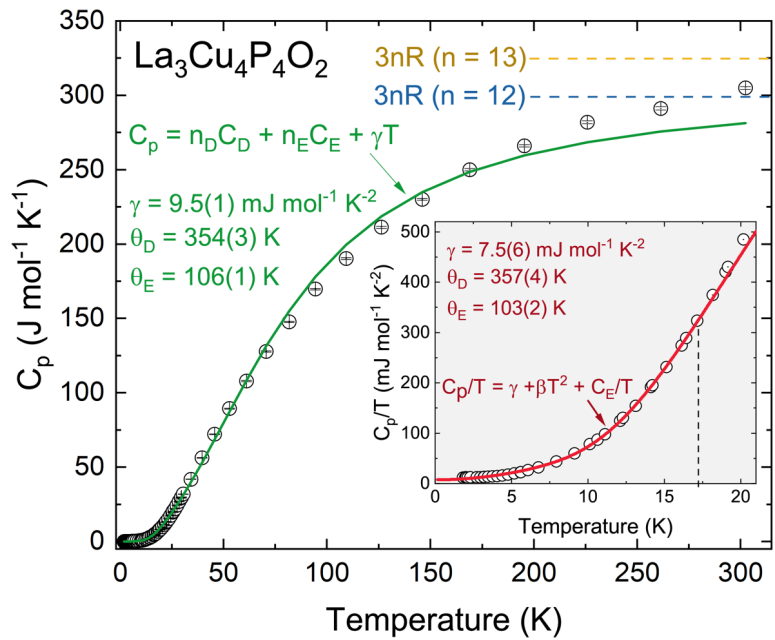

**Fig. 3.** Specific heat of $La_3Cu_4P_4O_2$ in the 1.9 K - 300 K temperature range. Green line is a fit to the data, which takes into account both Debye and Einstein contributions to $C_p$, with the total number of oscillators $n = n_D + n_E$. In the inset, $C_p/T$ vs T is presented in the low-temperature region, which was modeled as $C_p/T = \gamma + \beta T^2 + \frac{C_E}{T}$, where $C_E$ is the Einstein contribution (see text for details).

### Electrical resistivity

Temperature dependence of the electrical resistivity in the temperature range 1.9 - 300 K for $La_3Cu_4P_4O_2$ is displayed in Fig. 4a. The sample shows metallic behavior down to $T_{min} = 13.7$ K, with the value of residual resistivity ratio RRR = 3.6, similar to the observed (3.7) for the isostructural arsenide $La_3Cu_4As_4O_{2-\delta}$[6], while for $T < T_{min}$, there is an upturn in resistivity. The RRR of the $La_3Cu_4P_4O_2$ sample measured in[8] is 8.3, much larger than in the present study. While it could indicate reduced contamination of $Cu^{2+}$ ions from defects, in the discussion of magnetic susceptibility, we proposed an alternative explanation for the larger quantity of $Cu^{2+}$ ions in our sample compared to the sample studied in[8]. This alternative explanation also accounts for the lower RRR observed in our sample. Specifically, if there is a greater proportion of $Cu^{2+}$ ions, impurity scattering will be more pronounced-not due to actual impurities such as secondary phases-but because the equilibrium is shifted towards the left-hand side of the redox equation: $Cu^{2+} + P^{3-} \Leftrightarrow Cu^+ + P^{2-}$. While the change in the phosphorus oxidation state does not influence transport in a significant way, the presence of $Cu^{2+}$ instead of $Cu^+$ introduces additional scattering due to impurity spins, making resistivity larger. The $\rho(T)$ data were analyzed in two different temperature regions: below and over the minimum. For $T > T_{min}$, where the phonon scattering is the dominant mechanism, we employed the standard Bloch-Grüneisen relation for the temperature-dependent component: $\rho(T) = \rho_0 + \rho_{BG}$, where $\rho_{BG}(T) = 4R\left(\frac{\Theta_R}{T}\right)^5 \int_0^{\Theta_R/T} \frac{x^5 dx}{(e^x-1)(1-e^{-x})}$ is the Bloch-Grüneisen formula, $\rho_0$ is the residual resistivity, R is the gas constant and $\Theta_R$ is the calculated Debye temperature. The data were described well ($R^2 = 0.9988$) without taking into account other contributions to the electrical resistivity. The values extracted from the analysis are $\rho_0 = 97.1(2)$ $\mu\Omega$cm and $\Theta_R = 252(2)$ K, with Debye temperature reasonably close to the one calculated from the specific heat measurements.

Subsequently, we analyzed the low-temperature $T < T_{min}$ range. Low-temperature resistance of $La_3Cu_4P_4O_2$ under the applied magnetic field of $\mu_0 H = \pm 9$ T is shown in Fig. 4b, with the value of the resistance at the minimum subtracted to better visualize the effect of the applied magnetic field. What's interesting, the upturn below $T_{min}$ is suppressed under applied magnetic fields of $\pm 9$ T. In order to understand the origin of the observed $\rho_{min}$ in $La_3Cu_4P_4O_2$, $\rho(T)$ measurements at zero magnetic field below $T = 50$ K have been performed on $La_3Cu_4P_4O_{1.9}F_{0.1}$ and $La_3(Cu_3Ni)P_4O_2$ samples as well, presented in Fig. 4c. As seen from the figure, the upturn is sensitive to the substitution on the Cu-site; Ni-doping enhances the $T(\rho_{min})$ and decreases the magnitude of the upturn, while the F doping of the O site only slightly enhances the magnitude of the upturn below T = 13.7 K. In order to investigate the potential influence of grain boundary scattering on conduction electrons below T = 13.7 K, electrical resistivity measurements were performed on the $La_3Cu_4P_4O_2$ sample after annealing at 900°C overnight (see Fig. S5 in Supplementary Information). The results revealed that the annealing process strengthened the low-T increase in resistivity.

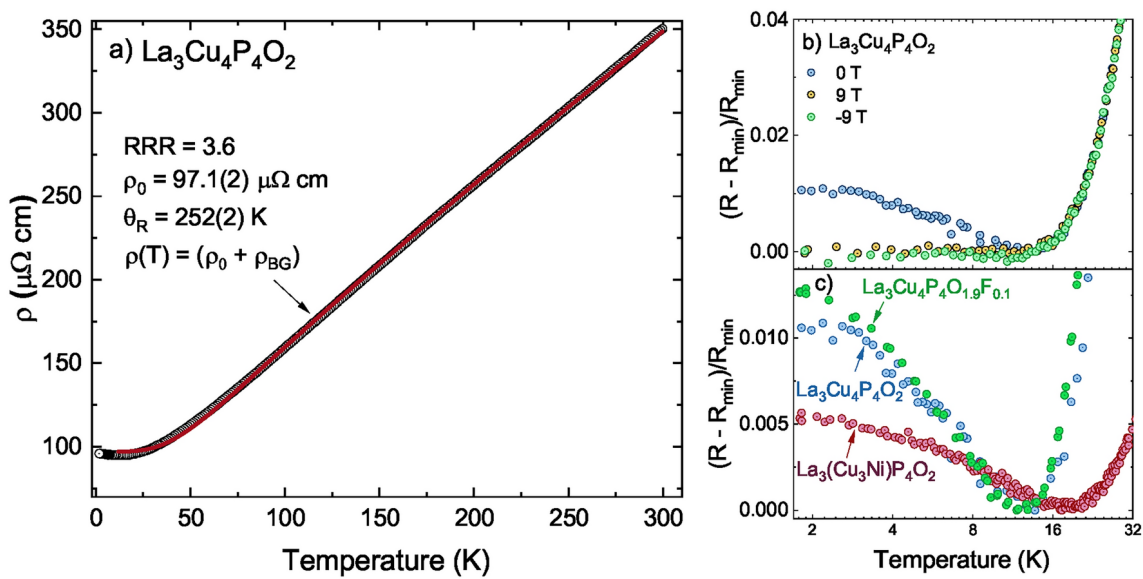

**Fig. 4**. (**a**) Resistivity of $La_3Cu_4P_4O_2$ in the 1.9 K - 300 K temperature range, modeled with the Bloch-Grüneisen formula (red line). (**b**) Low temperature resistivity of $La_3Cu_4P_4O_2$, presented as $(R - R_{min})/R_{min}$, showing the suppression of the minimum in the magnetic field. (**c**) Low temperature resistivity for doped $La_3Cu_4P_4O_2$ samples, presented as $(R - R_{min})/R_{min}$.

Minima in electrical resistivity were previously observed in several metallic materials, where they were attributed to the Kondo effect[46], magnetic superzone energy gap[47], and quantum corrections[48] to the resistivity which involve EEI and WL effects. The formation of magnetic superzone gaps in the ordered phase is not expected in the $La_3Cu_4P_4O_2$ system due to the lack long-range magnetic order down to 1.9 K. The electron - electron interaction (EEI) in a disordered system is characterized by a $\rho_{EEI}(T) \propto -T^{1/2}$ dependence that remains unaffected in the presence of the magnetic field[49]. As shown in Fig. 4b the upturn in the $\rho(T)$ is suppressed by the applied field which excludes the possibility that EEI dominates the low-temperature transport behavior of $La_3Cu_4P_4O_2$. The contribution to the resistivity due to the quantum interference effect in the weak localization (WL) scenario can be described by a $T^{-p/2}$ (p = 2, 3/2, 3 depending upon the scattering mechanism)[50], whereas for the Kondo effect, the resistivity should rise as $-\ln(T)$[9]. To elucidate the prevailing mechanism in $La_3Cu_4P_4O_2$, we plotted the low-temperature part of resistivity versus $T^{-p/2}$ and $\log(T)$ (Fig. S6.). Upon direct comparison of these plots, a conclusive determination regarding the scattering mechanism in the $La_3Cu_4P_4O_2$ could not be attained. However, based on the observation of $Cu^{2+}$ ions in the magnetic susceptibility measurements, we analyzed the data under the assumption that the resistivity upturn is caused by the Kondo effect.

For systems exhibiting the single-ion Kondo effect, the resistivity upturn at low temperatures can be modeled with a $\rho_{sd}$ term, which represents the resistivity arising from the scattering of conduction electrons by localized magnetic moments. This term is described by Hamann's expression: $\rho_{sd} = \frac{\rho_0}{2}\left(1 - \frac{\ln(T/T_K)}{\left[\ln^2(T/T_K)+S(S+1)\pi^2\right]^{\frac{1}{2}}}\right)$

[51], where $\rho_0$ is the unitary limit, $T_K$ denotes the Kondo temperature, and $S$ is the spin of the magnetic impurity. However, it is well known[52,53] that Hamann's expression cannot be applied at temperatures where $\rho(T)$ deviates from $-\ln T$ behavior, and yields small values of $S$, due to the inadequacy of the Nagaoka approximation[54] on which it is based[55,56]. This issue can be addressed by substituting $T_{eff} = \left(T^2 + T_W^2\right)^{1/2}$ for $T$[57] in the expression for $\rho_{sd}$, where $T_W$ refers to the energy scale of RKKY interactions, $k_B T_W$, between the localized impurities. In Fig. 5, we present the resistivity of $La_3Cu_4P_4O_2$, measured down to $T = 0.5$ K using a $^3$He insert. To analyze the experimental data, we utilized the procedure described in[52]. First, we modeled the resistivity in the 20-30 K temperature range with the expression (not shown): $\rho(T) = \rho_0 + AT^2$, where the first term is the residual resistivity and the second term represents Fermi liquid behavior. Then, fixing the value of $A$ as extracted from the first fit, we modeled the resistivity in the 5-25 K range with the formula $\rho(T) = \rho_0 + AT^2 + \rho_{sd}$ (cyan line), with $\rho_{sd}$ given by Hamann's expression as defined earlier. We did not attempt to perform the fit below $T = 5$ K as this is where $\rho(T)$ data deviates from $-\ln T$ behavior. For $S = 1/2$ ($Cu^{2+}$), we obtained $T_K = 14(1)$ K, which nicely reproduces the position of the resistivity minimum. Finally, substituting $T_{eff}$ for $T$ in the Hamann's

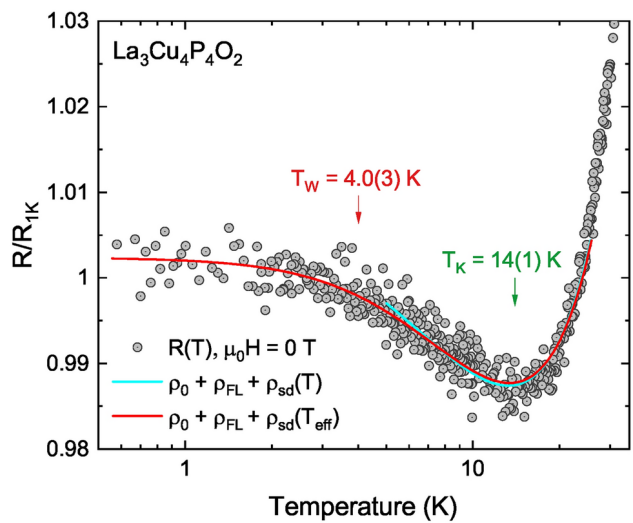

**Fig. 5.** Normalized low-temperature resistance of $La_3Cu_4P_4O_2$ vs T plotted in log scale. The temperature dependence of resistance below T = 25 K was modeled with two different expressions, taking into account scattering on lattice defects, Fermi liquid behavior and spin-dependent scattering (see text for details). $T_K$ and $T_W$ indicate Kondo temperature and temperature corresponding to the energy scale of RKKY interaction $k_B T_W$, respectively. The experimental data were obtained with the $^3$He PPMS insert.

expression, we performed the fit again, this time in the 0.5-25 K temperature range, with the value of $T_K$ and S fixed (red line). We obtained $T_W = 4.0(3)$ K, again nicely reproducing the temperature at which our $\rho(T)$ data deviate from linearity.

To gain more information about the underlying mechanism governing the resistivity upturn at low temperatures of the $La_3Cu_4P_4O_2$, magnetoresistance (MR) measurements at T = 2 K and T = 100 K were performed, below and above the resistivity minimum, respectively (Fig. S7 in Supplementary Information). The magnetoresistance is negative for $T < T_{min}$ and positive above, as expected for both WL and the Kondo effect. Although not explicitly addressed in[6], the isostructural $La_3Cu_4As_4O_{2-\delta}$ exhibits a similar, shallow resistivity minimum around 10 K[6] which persists against an applied magnetic field of 9 T. However, the magnetoresistance of $La_3Cu_4As_4O_{2-\delta}$ is positive even at T = 2 K, well below the $T_{min}$. Therefore, the mechanism behind the resistivity minimum might be different in these two isostructural compounds.

### Electronic strucure

Fig. 6a and 6b show the calculated band structure (BS) and density of electronic states (DOS) of Cu-3442, respectively. The DOS($E_F$) is dominated by Cu and P states, in agreement with the assumed picture of La-O layers playing the role of charge reservoir and separator for the electronically active Cu-P network. The calculated BS and DOS agree well with the results of spin-polarized DFT+U calculations included in the AFLOWlib database[58] (aflow:d0bb737e1b8210bb), and with the calculations included in the Materials Project (MP) database (mp-6309)[59,60]. Both results are consistent with copper atoms in a state close to filled-shell, nonmagnetic $Cu^+(d^{10})$. What's more, neglecting the Hubbard U in our calculations did not affect the electronic structure in any significant way. This observation aligns well with the calculated position (3 eV below the $E_F$) of most Cu-d states. The remaining states, positioned at the Fermi energy level, exhibit pronounced hybridization with phosphorus, leading to a substantial degree of delocalization and the observed metallic properties of Cu-3442.

For the Ni-3442 compound, BS and DOS are presented in Fig. 6c and 6d, respectively. Our results are in agreement with previous calculations performed with FLAPW-GGA method[61], however, employing the PAW method in this study allows for a detailed analysis of the chemical interactions within the structure. In contrast to Cu-3442, for the Ni-bearing compound the states at the Fermi energy level are predominantly of Ni-d character, with a reduced contribution from P-p orbitals. The shift of the Fermi level to the region of more localized states results in the enhancement of the calculated density of states: DOS($E_F$) = 5.2 states/f.u. for $La_3Ni_4P_4O_2$, compared to DOS($E_F$) = 3.7 states/f.u. for $La_3Cu_4P_4O_2$.

One of the most interesting features of the $ThCr_2Si_2$ and related structures is the ability of the Si-position (X) atoms to undergo dimerization, where the X - X pair distance serves as an additional degree of freedom that can be tuned with, e.g., pressure. Experimental observations with $^{31}$P NMR (in $La_3Cu_4P_4O_2$) and theoretical calculations (in $La_3Ni_4P_4O_2$)[61] have provided evidence of dimer formation in both compounds. To verify this result, in Fig. 7 we present the calculated electron localization function (ELF) for both Cu-3442 (a) and Ni-3442 (b) in the plane containing the $P_2$ dimers. The high value of ELF between two P atoms is a direct evidence for the formation of a covalent $\sigma$-bond resulting from the overlap of (mainly) P-$p_z$ orbitals. While the ELF between the P atoms is enhanced in both compounds, its magnitude is significantly reduced for the Ni-3442.

To gain insight into the chemical interactions present in both 3442 compounds, we performed calculations of the crystal orbital Hamilton populations (COHP) projected onto atomic states (pCOHP) and crystal orbital bond index (COBI). Additionally, we evaluated their values integrated up to the Fermi level (ICOHP and ICOBI, respectively), as shown in Fig. 8. The ICOHP value correlates with bond strength, with a more negative value indicating a stronger bond, while the value of ICOBI is the crystal orbital bond order. For the P - P interactions, the interlayer P - P distances for both $La_3Ni_4P_4O_2$ (2.58 Å) and Cu-3442 (2.27 Å) lie within the range of covalent interactions for phosphorus[44] (max. 2.8 Å). However, both ICOHP and ICOBI values are much larger for Cu-3442, in agreement with the larger electron localization between the P nuclei seen in Fig. 7. Consequently, the stronger P - P bond in $La_3Cu_4P_4O_2$ results from the increased overlap (shorter distance) between the P atoms. On the contrary, the Cu - P bond is significantly weaker compared to the Ni - P one. This is because the higher electron count for Cu means that more antibonding Cu - P states are filled, as seen in Fig. 8a. Put simply, in $La_3Cu_4P_4O_2$ the system prefers a strong P - P bond sacrificing the Cu - P one, while the opposite is true for $La_3Ni_4P_4O_2$.

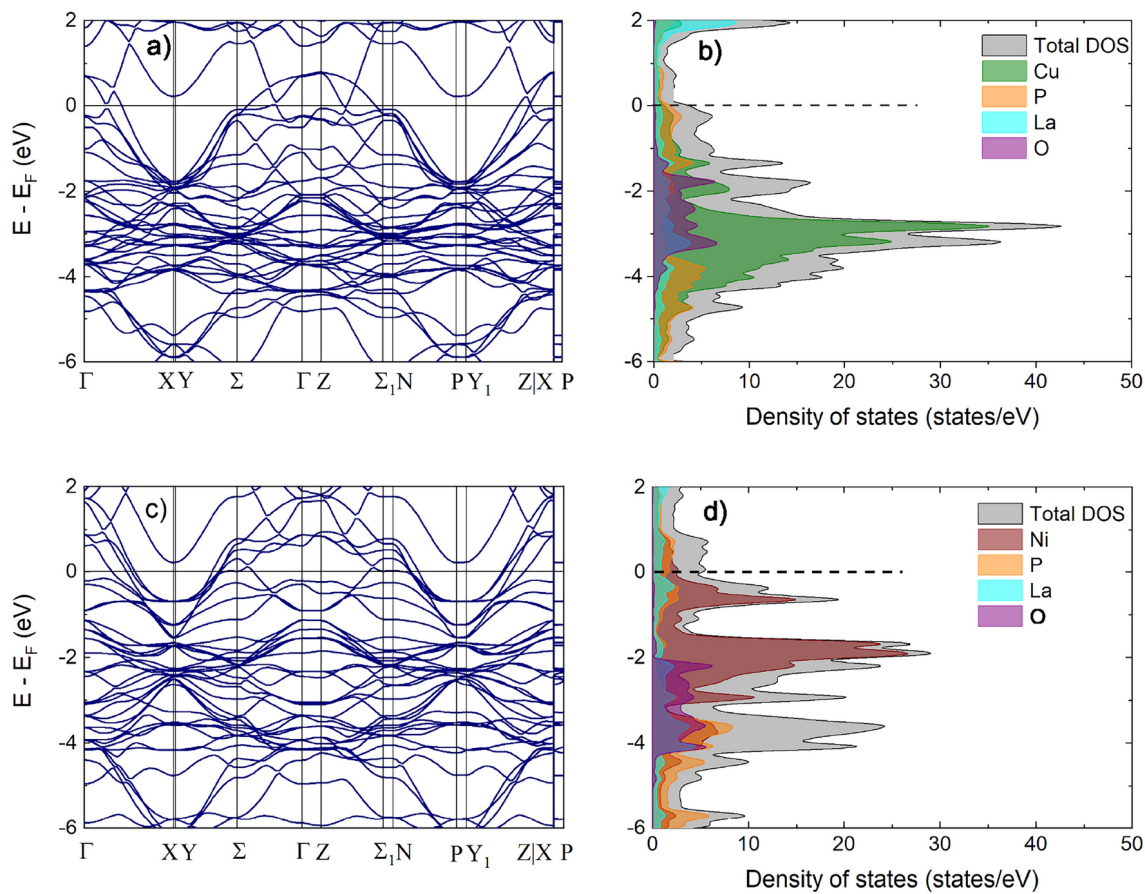

**Fig. 6.** (**a**) Electronic structure and (**b**) total (gray) as well as atom-projected (color-coded) density of states for $La_3Cu_4P_4O_2$. (**c**) and (**d**) the same for $La_3Ni_4P_4O_2$.

With the Ni-P subsystem more stable than the Cu-P one, the question arises as to why Ni-3442 manifests superconductivity, often concomitant with lattice instability[62–64]. In fact, the Ni-bearing 3442 compounds are generally less stable, as evidenced by both calculations ($La_3Ni_4P_4O_2$ is a metastable phase; see record no. mp-1210896[59,60]) and experimental observations: obtaining a single-phase Ni-3442 is more challenging than achieving the same for Cu-3442. With the Fermi level dominated by Ni and P states and the Ni - P interaction more stable, two possibilities could explain the compromised stability of $La_3Ni_4P_4O_2$. First, the stabilizing effect of Ni - P interaction can be outweighed by the destabilizing influence of the weaker P-P bond, due to the smaller overlap of the P atom orbitals, as we have shown in Fig. 8. Another one would be the T - T (T = Cu/Ni) interaction, and to check whether it's the case we plotted the Cu - Cu and Ni - Ni COHP in Fig. 9. Here, it's visible that both Cu - Cu and Ni - Ni interactions are very weak and practically nonbonding (ICOHP value close to zero), therefore it is expected that their effect on the stability of the crystal lattice is small. However, in $La_3Ni_4P_4O_2$ there is a peak of antibonding states of mainly $d_{xy}$ character just below the Fermi level. Therefore, in Ni-3442, both Ni-P interactions and, to a lesser extent, Ni-Ni interactions at the Fermi energy level are antibonding, which fact was previously linked to the occurrence of superconductivity[65,66]. In conclusion, the origin of the compromised stability of $La_3Ni_4P_4O_2$ results from the tendency of the system to prioritize the Ni - P interaction over the P-P one. This leads to the increased occupation of Ni-P antibonding states at Fermi energy level ($E_F$) and larger $DOS(E_F)$.

Finally, one might ask about the extent to which the comparison between the 3442 family and the iron pnictides is applicable. As noted in the introduction, these two families are closely related due to their similar crystal structures. However, the electronic structure calculation performed in this work point to the very different character of competing interactions present in the 3442 family when compared to iron pnictides. In the 122 structural family ($AEFe_2As_2$, AE - alkaline earth metal) the source of the instability is the competition between Fe - As covalence and Fe magnetism, with the Fe - As covalence effects strengthened by As - As interaction[67]. In the 3442 family, on the contrary, the Cu/Ni - P interaction is competing with the P - P one with no signs of magnetic order, whereas superconductivity occurs in $La_3Ni_4P_4O_2$ when the Ni - P bond wins over the P - P one. Moreover, the analysis presented in this manuscript points to the fact that the enhanced stability due to the stronger P - P bond as well as a smaller value of $DOS(E_F)$ of $La_3Cu_4P_4O_2$ may be the reason for the lack of superconductivity in this compound.

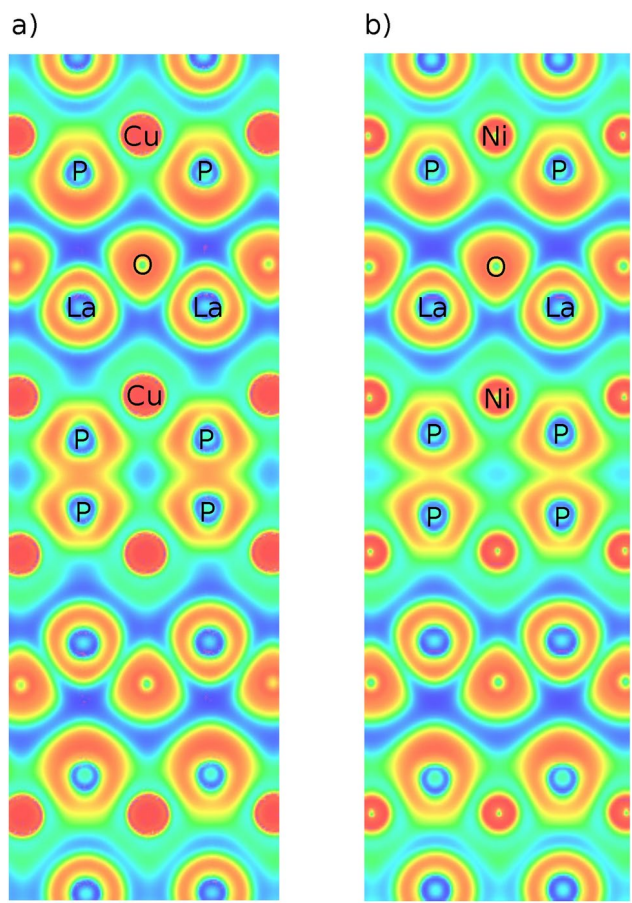

**Fig. 7**. (**a**) Calculated electron localization function (ELF) for $La_3Cu_4P_4O_2$ with the P - P interlayer bond shown. (**b**) The same for $La_3Ni_4P_4O_2$. Blue regions indicate low and red high values of the ELF, respectively.

## Summary
In conclusion, we have investigated the physical properties of $La_3Cu_4P_4O_2$ by means of magnetization, specific heat, and electrical resistivity measurements. No signs of magnetic ordering or superconductivity were observed down to the lowest temperatures obtained. DFT calculations and COHP analysis of the electronic structure data for $La_3Cu_4P_4O_2$ and isostructural superconducting $La_3Ni_4P_4O_2$ compounds have been performed. The results indicate that there is a competition between the Cu/Ni - P and P - P bonds. Specifically, it has been demonstrated that:

1. Enhancement of the Ni - P bond strength in $La_3Ni_4P_4O_2$ leads to an increased population of antibonding states at the Fermi energy level and larger $DOS(E_F)$, consequently inducing instability;
2. Conversely, a comparatively weaker Cu - P bond relative to the P - P bond in $La_3Cu_4P_4O_2$ contributes to the stability of the compound by reducing its $DOS(E_F)$, which might explain the lack of superconductivity in this compound.

We discussed the resistivity minimum at T = 13.7 K as resulting from the presence of $Cu^{2+}$ states necessary to obtain charge balance. Suppression of the resistivity minimum, together with the negative magnetoresistance at 2 K, could be consistently explained by the Kondo effect. However, the WL scenario could not be ruled out due to the noise level in the magnetoresistance measurements. High-quality single crystals of $La_3Cu_4P_4O_2$ are needed to further investigate the origin of the minimum in $\rho(T)$.

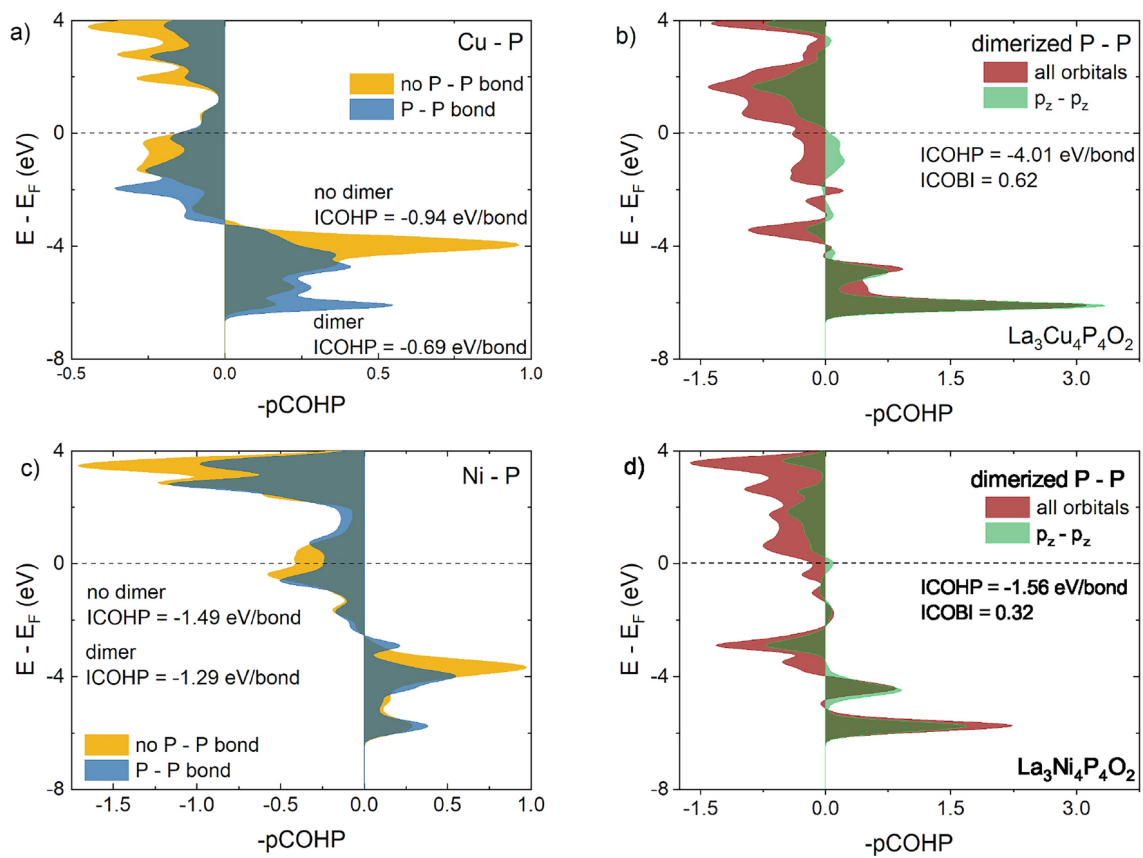

**Fig. 8.** (**a**) Partial crystal orbital Hamilton populations (pCOHP) for the Cu - P interaction with the nondimerized (yellow) and dimerized (blue) P atom in $La_3Cu_4P_4O_2$. (**b**) pCOHP for the dimerized P atoms in $La_3Cu_4P_4O_2$. The contribution of all (3s and 3p) orbitals is presented in brown, while the projected $p_z$ contribution is shown in green. (**c**) Same as in (**a**) but for the Ni - P interaction in $La_3Ni_4P_4O_2$. (**d**) Same as in (**c**) but for $La_3Ni_4P_4O_2$.

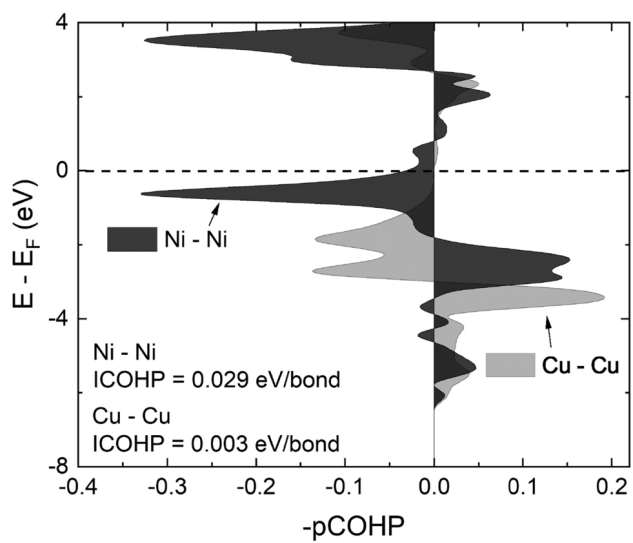

**Fig. 9.** Atom-projected -COHP (pCOHP) for Cu - Cu and Ni - Ni interactions present in $La_3Cu_4P_4O_2$ and $La_3Ni_4P_4O_2$, respectively.

## Data availability
The datasets used and/or analyzed during the current study are available from the corresponding author upon reasonable request.

## Acknowledgements
Sz. K. acknowledges the Technetium Talent Management Grants program (21/2022/IDUB/III.4.1/Tc) and D. Y. acknowledges the Nobelium Joining Gdańsk Tech Research Community project (38/2022/IDUB/I.1). Electrical resistance measurements using the $^3$He insert were performed at the Laboratory for Multiscale Materials Experiments at Paul Scherrer Institut.

## Author contributions
Sz. K. prepared the sample for measurements, analyzed the experimental data, conducted density functional theory (DFT) calculations, and drafted the manuscript. M. J. W. performed DFT calculations. D. Y. contributed extensively to the analysis of the experimental data. S. S. conducted electrical resistivity measurements using the $^3$He apparatus at the Laboratory for Multiscale Materials Experiments at the Paul Scherrer Institute. T. K. carried out physical property experiments and supervised the project. S. K., M. J. W., D. Y., S. S., and T. K. all contributed to the review of the manuscript.

## Declarations

### Competing interests
The authors declare no competing interests.

### Additional information
**Supplementary Information** The online version contains supplementary material available at https://doi.org/1 0.1038/s41598-025-89706-6.

**Correspondence** and requests for materials should be addressed to S.K. or T.K.

**Reprints and permissions information** is available at www.nature.com/reprints.

**Publisher's note** Springer Nature remains neutral with regard to jurisdictional claims in published maps and institutional affiliations.